# Local Controls for Large Assemblies of Nonlinear Elements

Michael Youssefmir and Bernardo A. Huberman

Dynamics of Computation Group
Xerox Palo Alto Research Center
Palo Alto, CA 94304

## Abstract

We introduce a set of local procedures that are capable of controlling distributed systems that exhibit complex dynamical behavior. These local controllers need only perturb local parameters and use local information about the state of the system. Besides eliminating the wiring and overhead needed for implementing a central controller, our procedure leads to desired states in short times. By resorting to a probabilistic dynamical argument we also show that there are critical values for the couplings among the nonlinear elements beyond which the local controllers do not achieve the desired final state.

## 1 Introduction

The recent trend towards embedding micromotors and actuators, along with computers, inside electromechanical structures [1], makes the control of these distributed systems both challenging and necessary. There are by now a number of adaptive structures in existence that modify their spatial configurations in response to rapidly changing environments, and novel applications are bound to appear in the near future. For example, attempts have been made at adaptively controlling conformal wings of aircraft so as to reduce drag, and a similar distributed approach has been advocated for controlling the surfaces of underwater vehicles [2, 3]. The current understanding of distributed systems has a long way to go [4]. As technology progresses, distributed control problems will become central in the implementation of electromechanical systems where the need for fast responses and reduced spatial complexity dictates the use of local, rather than global, control procedures, while still trying to achieve desired global behavior. This in turn opens up a number of questions, since relatively little is known about the distributed control of large assemblies of nonlinear elements.

Local control methods have already been shown to be quite powerful in the control of dynamic nonlinearities that may arise in distributed computer networks [5]. Since many of the same factors affecting network stability (delays in information, long feedback loops, bottlenecks in dealing with many elements in real time) conspire against the use of global controllers in the mechanical case, it is also desirable to have local controls acting on microelectromechanical structures. At the structural level, this can be achieved by having local controllers at each element site so that the system can reach the desired global state via a cooperative process. In doing so, one has to implement control mechanisms that take into account the dynamics of the elements as well as their spatial couplings, since disturbances in one actuator can propagate to adjacent ones in very short times.

Another interesting application of distributed control is to certain biological systems. For example, living cells control the geometrical and chemical structure of their membranes through local means that often involve small amounts of molecules. Such regulation schemes are composed of many interdependent components and operate in conditions that are far from equilibrium [6]. Since the understanding of such local mechanisms is considered a very challenging problem, the dynamics involved in controlling nonlinear assemblies might shed light on some of these issues as well.

In this paper we study the spatiotemporal control of a simple distributed dynamical system that can be representative of a large class of distributed nonlinear assemblies. These assemblies range from nonlinear oscillators to actuators, cilia and micromotors. Central to our approach is the assumption that the dynamics of each element of the system is easy to characterize around a particular orbit, where a linearized version of the dynamics can be used. However, no knowledge of the dynamics is needed outside this neighborhood, where the dynamics are highly nonlinear. For each element, there is a controller, and each controller is allowed to vary a local dynamical parameter within a given dynamic range. Using the linearized dynamics around the goal orbit, each controller initiates a parameter variation whenever the local system variable is close to the desired value. In this fashion, we manage to stabilize the global state of the system, and also show that the behavior of these controls is a statistical process in which competition between regions of controlled and uncontrolled behavior determines the eventual success of the control. This statistical approach to control in distributed nonlinear assemblies is in sharp contrast to the formal theory of linear control, based on linear algebraic methods, that has been developed within the last two decades (see [7] for a brief review).

In Section 2, we present a stroboscopic technique for controlling an array of nonlinear coupled elements, and local controllers are introduced to stabilize fixed points. Section 3 presents a statistical model for local control in large nonlinear assemblies which provides an explanation for the sharp transitions between controlled and uncontrolled behavior encountered in Section 2. Finally, we discuss the implications of these results for controlling nonlinear assemblies that will appear in the next generation of microelectromechanical systems.

## 2 The Model

The collective dynamics of coupled nonlinear oscillator networks has been studied extensively in many different systems, such as Josephson junction arrays [8] and coupled laser arrays [9] and a variety of biological systems [10]. Such networks can show a wealth of collective behaviors such as frequency locking, amplitude death, incoherence, partial locking, and even chaotic behavior. Indeed, there has been recent interest in the control of physical systems showing high dimensional chaotic behavior: oscillatory networks in a chaotic regime were considered in [11], experimental control of spatially extended globally coupled multimode laser was implemented in [12], and control of a system described by a partial differential equation was described in [13].

In what follows, we study an array of nonlinear dynamical elements that possess some coupling to each other. The nature of the coupling could be elastic or electromechanical, thus leading to a structure where any element can have an effect on the rest via local couplings. The whole system is required to be in a desired state, which for simplicity we will take to be a collective fixed point of the elements.

In order to keep such a distributed system operating within a desired regime, we consider a configuration, wherein local controllers are attached to each dynamical element. These controllers have access to the parameters determining the performance of each element and can vary them within an allowed range. This is done in order to avoid the problems of delays and information bottlenecks that exist when a global controller is used to regulate a spatially extended system.

To achieve global desired behavior with such local procedures, we propose a scheme whereby the dynamical state of the whole structure is sampled periodically by each controller. The schematic for the system is shown in Figure 1. This requires a timing mechanism at each controller. These local timing mechanisms can be synchronized via standard techniques (i.e. phase locked loops) that involve local interactions. By having synchronized timing mechanisms at each controller, the procedure then amounts to taking a Poincare section of the whole system, thus reducing the continuous time problem to one involving discrete dynamics on spatially distributed points. Each controller therefore has two components: a synchronization layer, that interacts with other controllers, and the actual control layer, that interacts only with the local system element. This leads to the control of a coupled map lattice of a type, whose dynamics have been used extensively to study spatiotemporal chaos [14].

The control algorithm that the local controllers use is based on recent work on stabilizing unstable orbits in systems exhibiting low dimensional chaos by making small perturbations to available system parameters at the point at which the system orbit has come sufficiently close to the unstable orbit [15, 16, 7].

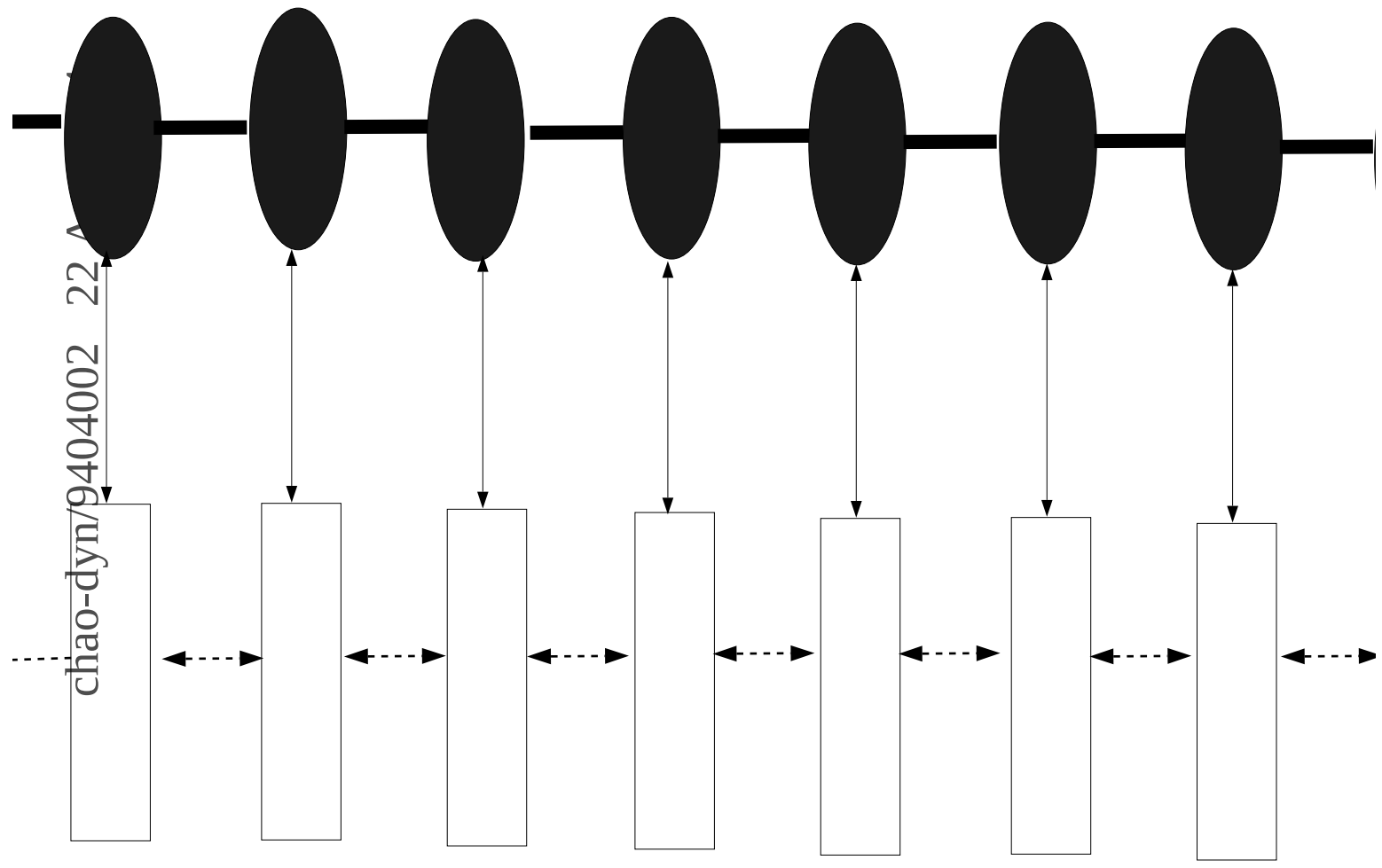

**Fig. 1.** Schematic showing the interconnectivities of the local controllers and the system variables. The local controllers (rectangles) have access to specific local parameters and information (vertical lines). They are also coupled to each other in order to synchronize local clocks (horizontal paths). System variables (circles) are manipulated by the local controllers and are coupled to each other.

The generic form for the coupled map lattices that results from stroboscopically sampling the distributed system is given by

$$x_{n+1}^{(i)} = F\left(x_n^{(i)}, a^{(i)}\right) + g^{(i)}\left(x^{(1)}, ..., x^{(N)}\right), \tag{1}$$

where $n$ is the time index, $i$ is the spatial index on the lattice of size $N$ ($i$=1,...,N),

$$F\left(x, a^{(i)}\right) \tag{2}$$

is a function describing the local dynamics of the nonlinear elements (in the examples below this will typically be the logistic map), $a^{(i)}$ is the local control variable at lattice site $i$, and $g^{(i)}$ is a function describing the coupling of the $i$th lattice element to the other lattice sites.

The object of the controllers is to stabilize the lattice at a particular unstable fixed point of the lattice map. The control of the lattice is to be implemented through local changes in the parameters $a^{(i)}$. Typically, the controllers will be limited to varying these control parameters in a fixed range around some base value, $a$, so that

$$\left|a^{(i)} - a\right| < \delta_{max}. \tag{3}$$

The motivation for such a restriction is two fold: first, such a range might be the only experimentally accessible one; and second, it might not be possible to make rapid adjustments in the parameters for too large a dynamic range. The control method that we used here is thus a parametric control of the variables $a$; as opposed to a feedback control of the actual system state variables. Moreover, we assume that the linearized dynamics around the goal state are known and that these linearized dynamics can indeed be stabilized with only the information available at the specific lattice point.

## Local Control I

We begin by considering an array of oscillators with nearest neighbor couplings, such that their Poincare sections give rise to the following lattice:

$$x_{n+1}^{(i)} = a^{(i)} x_n^{(i)} \left(1 - x_n^{(i)}\right) + \frac{\epsilon}{2}\left(x_n^{(i+1)} + x_n^{(i-1)} - 2x_n^{(i)}\right) \tag{4}$$

with $0 < \epsilon < 1$. For certain values of $a^{(i)}$, this represents highly nonlinear behavior locally along with a coupling that tends to draw each oscillator to the average of its nearest neighbors.

The dynamics at the lattice points are further constrained to stay in the unit interval by replacing values obtained from Eq. 4 outside the unit interval by a corresponding value in the unit interval. This is obtained by reflecting the value around 0 or 1 depending on which side of the unit interval the value of the function at that iteration is.

The aim of the controllers is to stabilize the homogenous state, given by $x_*^{(i)} = 1 - \frac{1}{a}$, which is otherwise unstable. A linear stability analysis about this fixed point shows that the variation in $a^{(i)}$, that causes each $x^{(i)}$ to fall onto the unstable fixed point, depends not only on the lattice site at the $i$th lattice point, but also on the lattice values of the nearest neighbors. Assuming that the system state is close enough to the desired fixed point, the parameter variations that guarantee this fastest approach to the fixed point are given by

$$a^{(i)} = a\left(1 + (a-2)\frac{\left(x^{(i)} - x_0\right)}{x_0} + \frac{\epsilon}{2x_0}\left(x^{(i+1)} + x^{(i-1)} - 2x_0\right)\right). \tag{5}$$

In the spirit of locality, these controllers might be implemented in such a way that the last term in Eq. 5 is dropped. In what follows the parameter variations used are then,

$$a^{(i)} = a\left(1 + (a-2)\frac{\left(x^{(i)} - x_0\right)}{x_0}\right). \tag{6}$$

A linear stability analysis of the dynamics, with the controllers given by Eq. 6, shows that, under these conditions, the homogenous fixed point of the lattice is linearly stable for values of $\epsilon$ that are not too big. Specifically, linear stability of the controlled lattice is guaranteed if the eigenvalues of the Lyapunov matrix, which, for the system described by Eqs. 4 and 6, is given by

$$\begin{pmatrix} \frac{\epsilon}{2} & -\epsilon & \frac{\epsilon}{2} & 0 & \dots & \dots & 0 \\ 0 & \frac{\epsilon}{2} & -\epsilon & \ddots & \ddots & \ddots & \vdots \\ \vdots & \ddots & \ddots & \ddots & \ddots & \ddots & \vdots \\ \vdots & \ddots & \ddots & \ddots & \ddots & \ddots & \vdots \\ 0 & \ddots & \ddots & \ddots & \ddots & \ddots & 0 \\ \frac{\epsilon}{2} & 0 & \ddots & \ddots & 0 & \frac{\epsilon}{2} & -\epsilon \\ -\epsilon & \frac{\epsilon}{2} & 0 & \dots & \dots & 0 & \frac{\epsilon}{2} \end{pmatrix}, \tag{7}$$

are all of absolute value less than 1. This occurs for $\epsilon < 0.5$. The controller behavior given by Eq. 6 is taken to be fully local in the sense that the parametric variation is initiated at each lattice point, independent of the dynamics at every other lattice point. The variation in the control parameter is made whenever the lattice value $x^{(i)}$ falls within a region, from here on called the region of controllability, and such that the associated control parameter $a^{(i)}$ satisfies

$$\left| a^{(i)} - a \right| < \delta_{max}. \tag{8}$$

If $x^{(i)}$ falls outside this region no parameter variation is performed and the controller is off. In all then the dynamics are given by,

$$x_{n+1}^{(i)} = a\left(1 + (a-2)\frac{\left(x^{(i)} - x_0\right)}{x_0}\Theta\left(x^{(i)}\right)\right)x_n^{(i)}\left(1 - x_n^{(i)}\right) + \frac{\epsilon}{2}\left(x_n^{(i+1)} + x_n^{(i-1)} - 2x_n^{(i)}\right), \tag{9}$$

where

$$\Theta\left(x^{(i)}\right) = \begin{cases} 1 & \left|(a-2)\frac{\left(x^{(i)}-x_0\right)}{x_0}\right| \le \delta_{max} \\ 0 & \left|(a-2)\frac{\left(x^{(i)}-x_0\right)}{x_0}\right| > \delta_{max} \end{cases}. \tag{10}$$

An example of the behavior of the lattice controllers is shown in figure 2 for $\delta_{max} = 0.9$ and $a = 3.9$. and values of $\epsilon$=0.205 and $\epsilon$=0.208 for 1000 iterations starting at the iteration 2400.

In the case of $\epsilon$=0.205, the goal state is typically achieved within about 10,000 iterations; while in the case of $\epsilon$=0.208, not even 100,000 iterations are enough to achieve

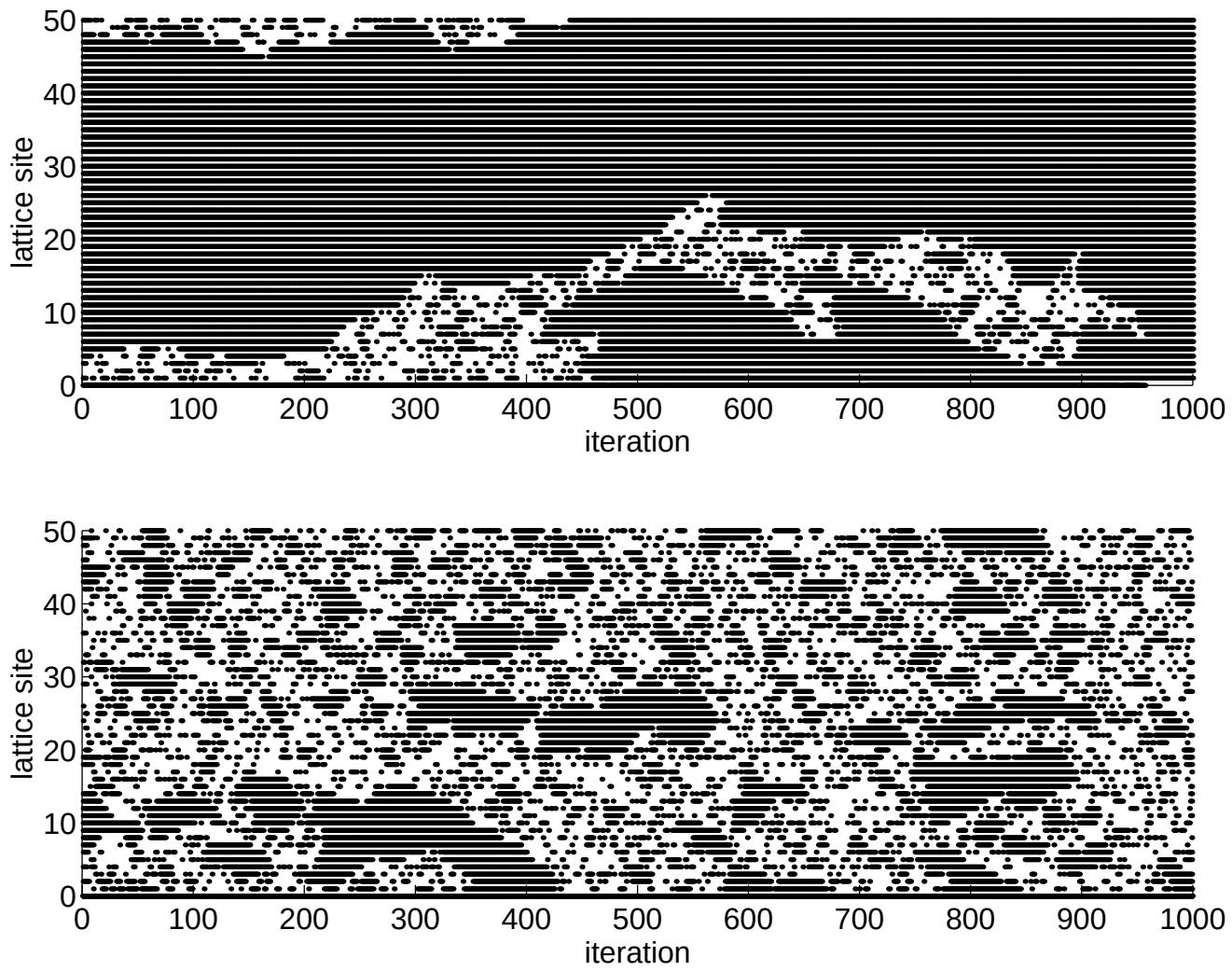

**Fig. 2.** Vertical axis represents lattice sites and horizontal axis represents number of iterations from random initial conditions for the lattice given by (eq. 4) and with parameters $a = 3.9$ and $\delta_{\max} = 0.9$. A dark pixel represents the lattice point being in the control region while a blank pixel represents the lattice point being outside the control region. The upper figure shows the lattice with $\epsilon$=0.205 for the 1000 iterations starting at the 2400th iteration. The bottom figure shows the lattice with $\epsilon$=0.208 for the same iterations.

the desired state. Figure 3 shows the number of iterations needed to achieve control at the desired state for different values of $\epsilon$ with $\delta_{max} = 0.9$ for 50 different initial conditions. If control was not achieved within 200,000 iterations a value of 200,000 was recorded. Of course, this hides the fact, that will be explored below, that the system will never reach the fixed point for certain parameter values. For $\epsilon$=0.205, the lattice relaxes to the desired state within the 200,000 iterations; however, for $\epsilon$=0.207, the controllers mostly fail to achieve the fixed point within 200,000 iterations; and, for $\epsilon$=0.208 and above control could not be reached for any of the trials within the 200,000 iterations. Indeed, it is clear that the overall success of the controllers is quite sensitive to the couplings in the lattice and to the dynamic range of the controllers

It is important to contrast this local implementation with a global implementation of a controller for the lattice. A global control could, for example, ramp down all the control parameters to a region in which the lattice is naturally attracted to the homogenous fixed point. The dynamic range for $a^{(i)}$ might indeed allow this possibility. Once this is done, the parameters could slowly be ramped up in such a way as to keep the state of the lattice at its fixed point. There are two problems with such a scenario: first, such a central controller is not very robust to external fluctuations in that if one of the lattice points is kicked out of the stable regime, the entire lattice has to be modified (in some sense shut down) in order to correct the situation; and second, delays in the communication of the state of the local variables to the central controller can be substantial. Such delays can cause the parametric controller to lose stability and make the central controller dysfunctional.

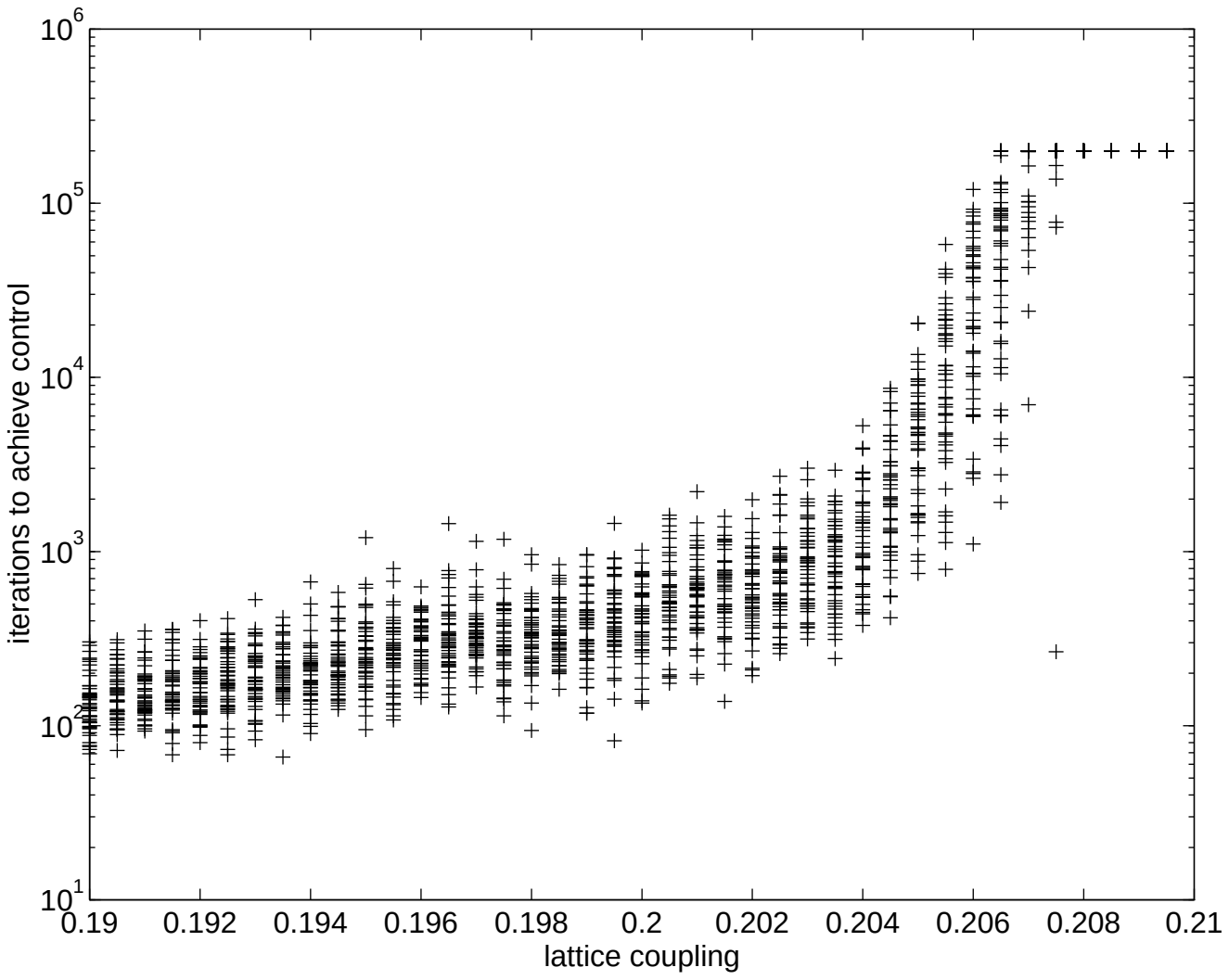

**Fig. 3.** Data points for time necessary to achieve control of the lattice given by Eq. 4. Vertical axis shows the time necessary to achieve full control for a given value of $\epsilon$ on the horizontal axis. For each $\epsilon$, 50 runs of maximum length 200,000 were used. For runs that did not reach control in the given 200,000 iterations, a value of 200,000 was recorded.

Another global control might take the equations of motion and calculate an optimal control trajectory to the desired state by using the nonlinearities in those equations. Such an implementation would require a great deal more knowledge of the system equations, as well greater computational resources, than the solution presented here.

## Local Control II

Next we consider a coupling map lattice with a diffusive nearest neighbor coupling that has been studied extensively [14, 17],

$$x_{n+1}^{(i)} = (1-\epsilon)f\left(x_n^{(i)}\right) + \frac{\epsilon}{2}f\left(x_n^{(i-1)}\right) + \frac{\epsilon}{2}f\left(x_n^{(i+1)}\right), \tag{11}$$

where

$$f\left(x_n^{(i)}, a^{(i)}\right) = a^{(i)}x^{(i)}\left(1 - x^{(i)}\right), \tag{12}$$

and $0 < \epsilon < 1$. Lattice point dynamics are as before constrained to stay in the unit interval by replacing values obtained from the function $f(x)$ outside the unit interval by the corresponding value in the unit interval obtained by reflecting the value around 0 or 1. It should be noted that this same map was used by [17] wherein in a feedback to the actual state variables was used to control and eliminate fluctuations, chaotic and otherwise. Such feedbacks are equivalent to parametric variations when near the unstable fixed point.

Our aim here is again to stabilize the homogenous state, i.e. that in which all elements are in the same dynamical state. It is given by $x_*^{(i)} = 1 - \frac{1}{a}$. A simple calculation shows that the linearized system about the fixed point can be stabilized by the control algorithm

$$a^{(i)} = a\left(1 + c^{(i)}\left(x^{(i)} - x_0\right)\right), \tag{13}$$

where the constants $c^{(i)}$ can take on a range of values to achieve stability about the fixed point. Indeed a linear stability analysis around the fixed point gives the condition that

$$\frac{a-3}{x_0} < c^{(i)} < \frac{a-1}{x_0}. \tag{14}$$

Control is again initiated at each lattice site via the feedback Eq. 13 whenever the lattice value $x^{(i)}$ falls within the controllability region. In the simulations below, the dynamic range of the local control parameters is given by $\delta_{max} = 0.9$. The value $c^{(i)} = \frac{a-2}{x_0}$ is used, and control is initiated at a lattice site only if the modification to the control parameter falls within the allowable range. This value for $c^{(i)}$ is optimal in the sense that, in the linearized dynamics about the fixed point, it guarantees immediate relaxation to the fixed point. When the use of this optimal parameter variation is disallowed by Eq. 3, it is still possible that a less than optimal adjustment can be made. Such adjustments will not be taken advantage of below.

The fixed point under this control mechanism is linearly stable for all values of $\epsilon$. At each iteration, the lattice is first updated, then the controllers are allowed to act on each control variable. Figure 4 shows the resulting dynamics on a lattice of size 50 for the uncontrolled and controlled lattices with parameter values $a = 3.9 \text{ and } \epsilon = 0.35$. As can be seen, the controllers have indeed controlled the fluctuations in the lattice, but the lattice has been attracted to a different dynamical state than the one to be stabilized. It is not hard to see the origin of these patterns: Specific lattice sites might, for example, have both the neighbors on either side locked at the fixed point, while they remain uncontrolled because their stable periodic orbits are outside the control region. This is the origin of the extraneous states in figure 4, and in a local control implementation, it is important to realize that such anomalous stabilizations might occur.

It is also important to note that such spatial patterns may not be very representative of the behavior of continuous time systems. As shown in [18], such patterns disappear when the lattice updates are made asynchronously. To illustrate this effect, the same controllers were “turned off” with a 10% probability when the corresponding lattice point was in the control region. The final plot in figure 4 shows how this modification eliminates the “novel” spatial patterns that might arise in a purely synchronous updating.

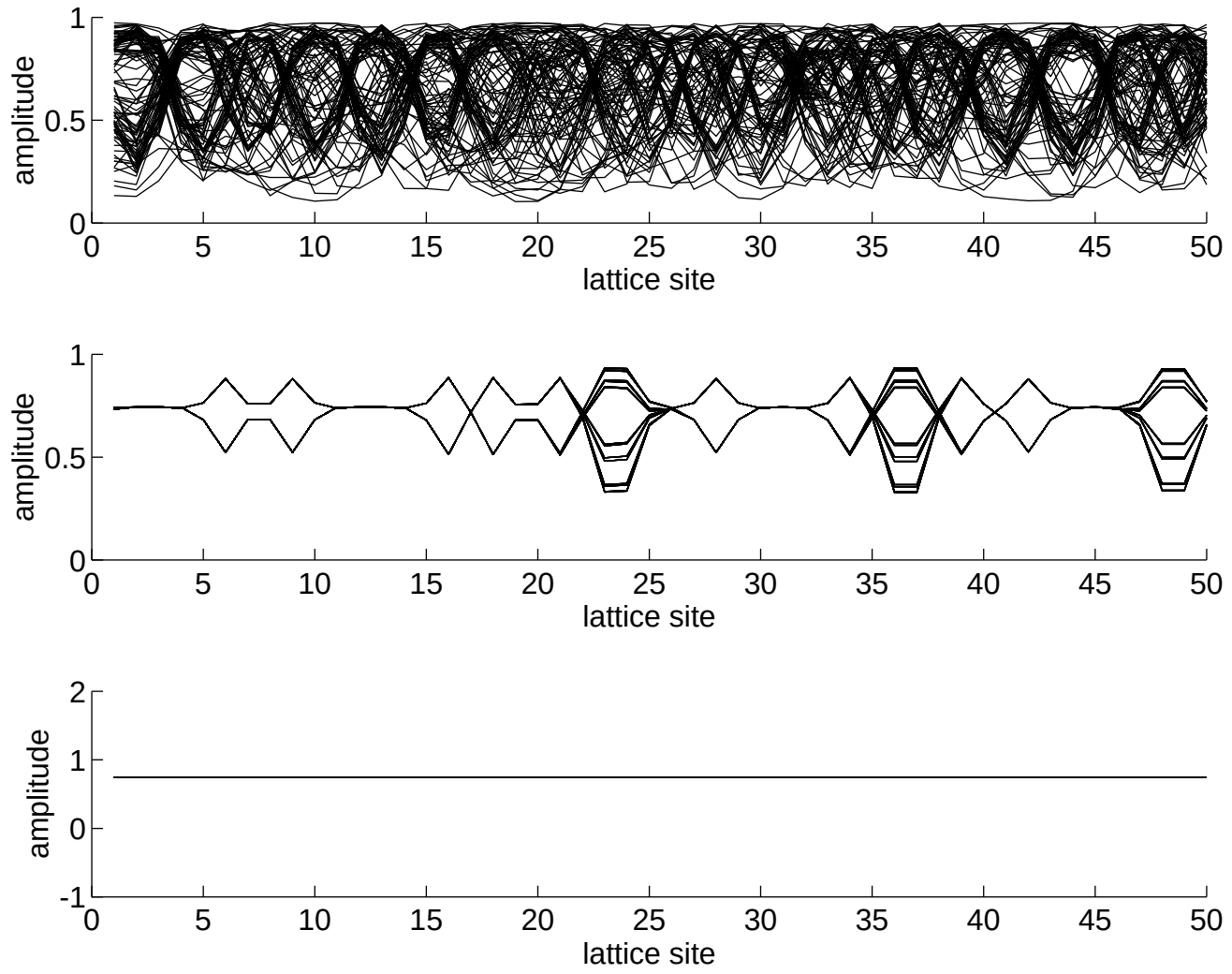

**Fig. 4.** Space amplitude plots for 100 superimposed iterations for a lattice of size 50 given by Eq. 11. The lattice parameters are $a = 3.9$ , $\epsilon = 0.35$, and $\delta_{\mathrm{max}} = 0.9$. Horizontal axis shows lattice site and vertical shows site values. First plot shows the uncontrolled lattice. Second plot shows controlled lattice. Third plot shows controlled lattice with controllers having a 10% probability of being turned off to mimic asynchronicity when lattice site is in control region.

We should point out that it is also possible to apply these control strategies to orbits with period greater than one (ie. limit cycles instead of fixed points). Again, the controllers are initiated if the corresponding lattice point falls within a certain critical region around one of the target orbit points; however, each individual controller simply tries to stabilize itself and no information about the state of other lattice points is used. For the situation in figure 5, for example, the period 2 orbit for $a = 3.9$ and $\epsilon = 0.5$ is reached simply by allowing lattice points to execute the control strategy using simply the standard linear control strategy whenever the orbit at the specific lattice falls in a critical region, defined by the constrained $\delta_{max} = 0.9$, around either of the period 2 orbit points. It should be stressed again that the controllers again need no information about the state of other lattice points. The final phase of the period 2 orbit is determined by competition between regions of different phases. Indeed, control of periodic orbits with periods higher than 1, in most cases settle into regimes with large regions of in phase behavior separated by relatively small regions of kinks. Such states tend to disappear as $\delta_{max}$ or $\epsilon$ are increased.

Once again, the implementation of local controls does not guarantee global stabilization at the fixed point. Indeed, the success of the lattice control is quite sensitive to the values of $a_0$, $\epsilon$, and $\delta_{max}$. The first plot in figure 6 shows a situation ($a = 3.9$, $\epsilon = 0.517$, and $\delta_{\mathrm{max}} = 0.9$ ) in which control of the lattice period one orbit is not achieved within 30000 iterations. Although, for some initial conditions control is achieved, typically the lattice does not stabilize at the desired state. Large regions of controlled lattice points exist but are then slowly destroyed by the disordering behavior of moving dynamical "defects" to neighboring lattice points. However, by making a small adjustment in the value of the diffusive coupling $\epsilon$ (ie. $\epsilon = 0.515$) , the controllers bring the lattice to the desired state, as shown in the second part of the figure.

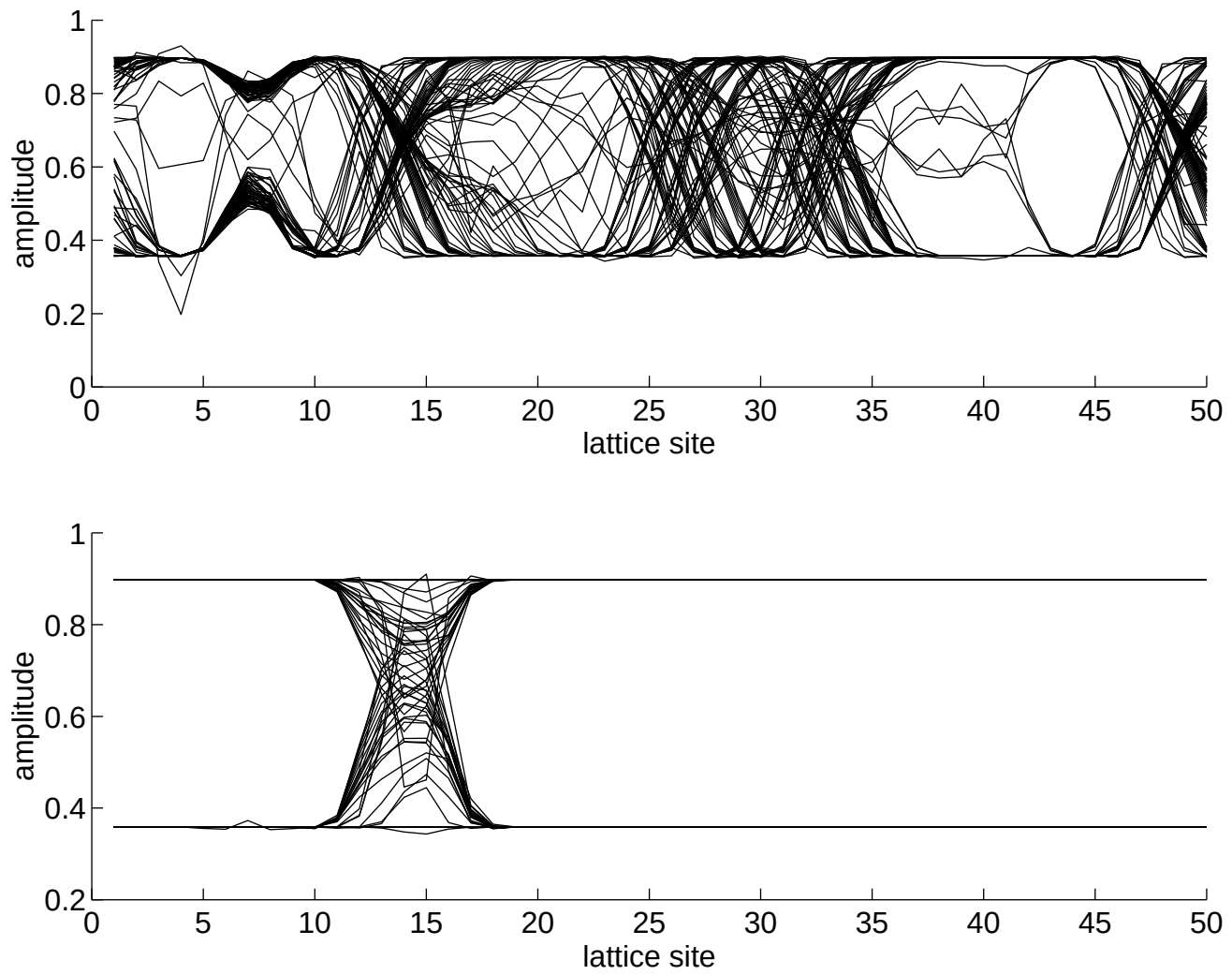

**Fig. 5.** Stabilization of the period 2 orbit in the lattice Eq. 11, for $a = 3.9$ , $\epsilon = 0.35$, and $\delta_{\mathbf{max}} = 0.9$. The initial fluctuations (upper) superimposed on each other and (bottom) the final iterations before full period 2 behavior

Once again, it is not clear to what extent such dynamics can be observed in a continuous time implementation of a spatially distributed system. The lattice control breaks down because small regions of uncontrolled lattice points which cannot be controlled by the local controllers propagate along the lattice. Evidently for values of $\epsilon$ slightly smaller such defective regions are no longer stable and are eventually stopped by the controllers.

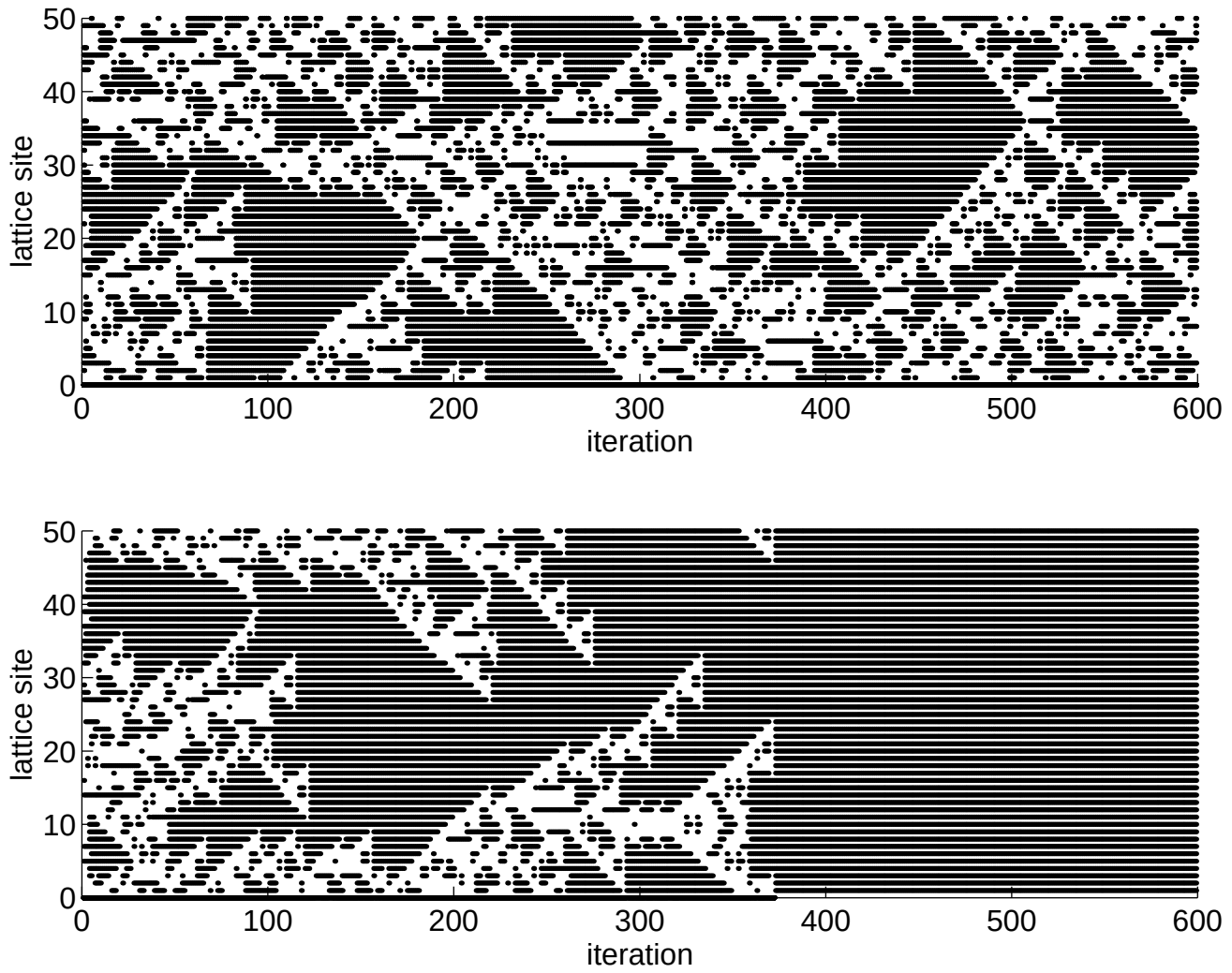

**Fig. 6.** Vertical axis represents lattice sites and horizontal axis represents number of iterations from random initial conditions for the lattice given by eq. 11 with parameters $a = 3.9$ and $\delta_{\mathrm{max}} = 0.9$. A dark pixel represents the lattice point being in the control region while a blank pixel represents the lattice point being outside the control region. The upper figure shows the lattice with $\epsilon$=0.517 for 600 iterations starting from random initial conditions (typically the lattice settles into a state in which lattice points are all controlled except for moving defects) The lower figure shows the lattice with $\epsilon$=0.515 for the first 600 iterations (full control is typically achieved).

# 3 The Dynamics of Local Controls

We now generalize the above models into a theoretically more tractable framework. In order to do so we make the following assumptions: first, that the dynamics of the system are such that an uncontrolled lattice point traces an orbit that is essentially random; second, that an uncontrolled lattice point has an associated nonzero probability, denoted $P(U \to C)$, of falling onto the control region; and, finally, that a controlled lattice point itself has an associated probability, denoted $P(C \to U)$, of being kicked out of the dynamic range of the controller and becoming uncontrolled. With these assumptions,

$$P(U \to C) = \alpha, \tag{15}$$

where $\alpha$ is a constant. The probability $P(C \to U)$, will generally depend on the probability that the lattice point is coupled to an uncontrolled lattice point, and that this uncontrolled lattice point is in a region that will knock the controlled lattice point out of the control region. This probability will increases with higher coupling and decreases with an increase in the controller dynamic range. Generally we then take,

$$P(C \to U) = \beta f_u. \tag{16}$$

where $f_u$ is the fraction of lattice elements that are uncontrolled, and $\beta$ describes the probability that an uncontrolled lattice point is in a region that tends to disrupt a controlled lattice point through the lattice coupling. Using a mean field approximation to the dynamics, the number of uncontrolled elements that tend to be controlled at each iteration is $\alpha N_u$; and the number of controlled elements that fall out of control at each iteration is $\beta f_u N_c$. $N_c$ and $N_u$ are the number of controlled and uncontrolled elements respectively. Thus the fraction, $f_c(i+1)$, of controlled elements at the $(i+1)$th time step is related to the fraction at the previous time step by,

$$\begin{aligned} f_c(i+1) &= f_c(i) + \alpha(1 - f_c) - \beta(1 - f_c) f_c \\ &= \beta f_c(i)^2 + (1 - \beta - \alpha) f_c(i) + \alpha. \end{aligned} \tag{17}$$

This equation has a stable fixed point at $f_c = 1$ for $\left(\frac{\alpha}{\beta}\right) > 1$, which becomes unstable for $\left(\frac{\alpha}{\beta}\right) < 1$. There is then a new stable fixed point at

$$f_c = \frac{\alpha}{\beta}. \tag{18}$$

As $\left(\frac{\alpha}{\beta}\right) \to 1$ from above, the average time for the approach to the fully controlled state diverges as,

$$\tau_{control} \sim \frac{1}{\log\left(1-\alpha+\beta\right)} \sim \frac{1}{1-\frac{\alpha}{\beta}}. \tag{19}$$

Finally, for $\left(\frac{\alpha}{\beta}\right) < 1$ the full control of the system is not possible, and the system falls into a state with an average number of controlled elements that is less than one.

For the lattice map given by Eq. 4, since the probability distribution for an uncontrolled lattice point is smooth, we expect that, for small $\delta_{max}$, $\alpha$ will typically scale linearly with the dynamic range of the controller, $\alpha \sim \delta_{max}$ and will to first order be independent of $\epsilon$. Furthermore, since lattice points are coupled linearly to neighboring lattice points, we expect that $\beta$ will typically scale up with the coupling,

$$\beta \sim \epsilon. \tag{20}$$

Together with Eq. 20, this predicts that as the strength of the lattice coupling, $\epsilon$, in the lattice map of Eq. 4 is increased from zero, the number of controlled elements remains constant up to a critical value, falling off afterwards inversely with the strength of the lattice coupling.

In order to test these ideas we ran a computer experiment for a lattice of size 1000, varying the lattice coupling constant from 0.19 up to 0.23. As figure 7 shows there is indeed a very sharp drop in the ability to control these elements once a critical value of the coupling constant is reached. It is worth remarking that this mean field theory does not predict the observed value of the critical coupling, nor does it explain the actual crossover to the long tail of the transition.

It is also possible to extend the above analysis to a continuous time system. In this case, the probabilities P(C→U) and P(U→C) can be thought of as probabilities to change states in a sufficiently short time, Δt. Formally, they become

$$P(U \to C) = \alpha \Delta \mathrm{t} \tag{21}$$

and

$$P(C \to U) = \beta f_U \Delta \mathrm{t}. \tag{22}$$

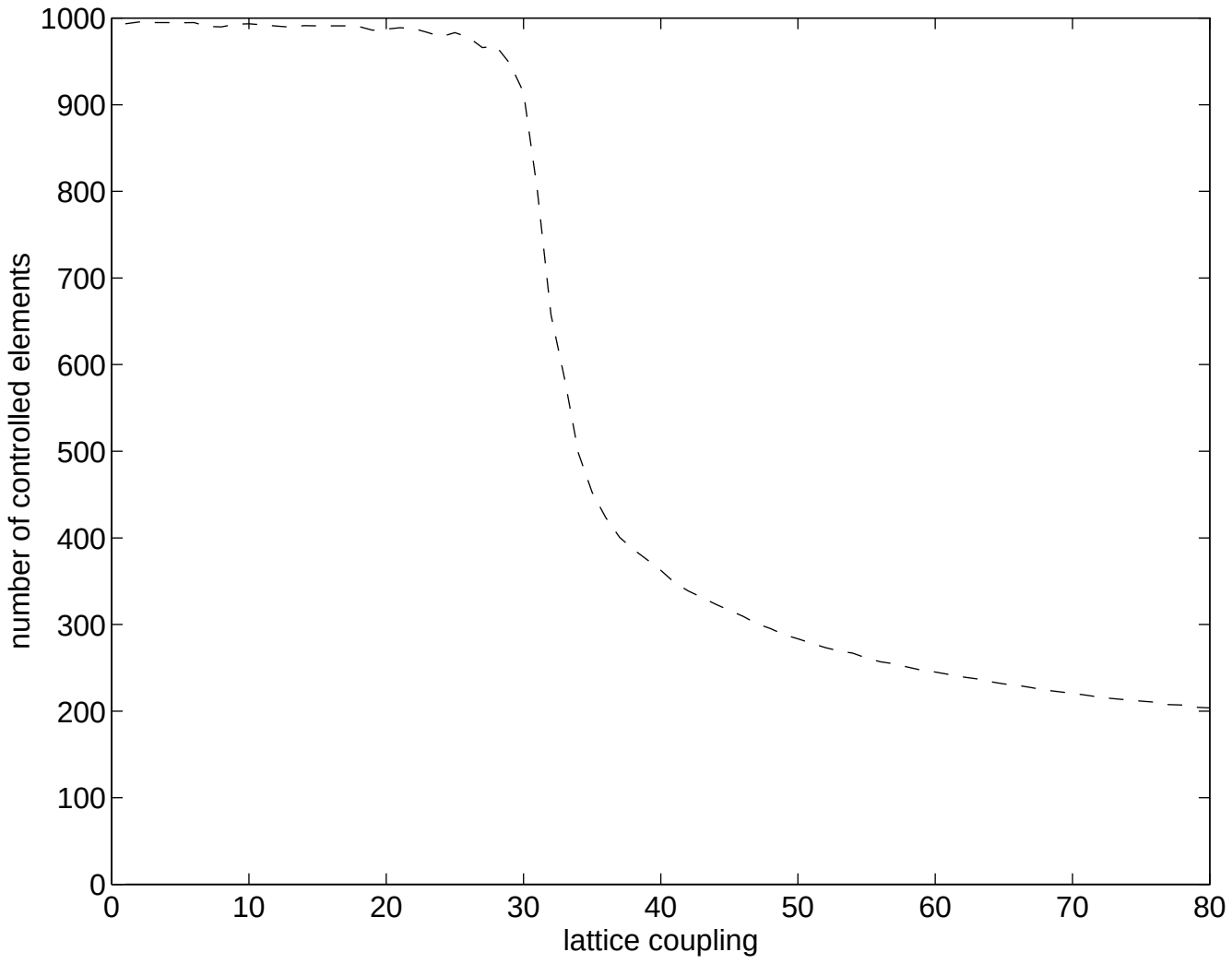

**Fig. 7.** The average number of controlled lattice points as a function of $\epsilon$ for the lattice dynamics given by 4 with size 1000 and with $a = 3.9 \;\; \text{and} \;\; \delta_{\max} = 0.9.$.

Using a master equation, the probability, $P(N_c.t)$, that $N_c$ elements are controlled satisfies,

$$\frac{dP(N_c)}{dt} = P(N_c - 1)[N - N_c + 1]\alpha + P(N_c + 1)[(N_c + 1)(1 - f_c)]\beta \\ + P(N_c)[-(N - N_c)\alpha - N_c\beta(1 - f_c)]. \tag{23}$$

Multiplying this equation by $N_c$ and taking the mean field limit ( ie. $\langle N_c^2 \rangle = \langle N_c \rangle^2$), we obtain an equation for $f_c$ of the form,

$$\frac{\mathrm{d}f_c}{\mathrm{dt}} = \beta f_c^2 - (\beta + \alpha) f_c + \alpha. \tag{24}$$

This equation can be solved to give

$$f_c(t) = \frac{\left(\frac{\alpha}{\beta}\right) C \exp\left[\beta\left(1 - \left(\frac{\alpha}{\beta}\right)\right)t\right] - 1}{C \exp\left[\beta\left(1 - \left(\frac{\alpha}{\beta}\right)\right)t\right] - 1}, \tag{25}$$

where $C = \left(\frac{\left(\frac{\alpha}{\beta}\right) - f(0)}{1 - f(0)}\right)$, and it produces the same fixed points as the discrete version that we analyzed above,

$$f_c(t \to \infty) = \begin{cases} \frac{\alpha}{\beta} & \left(\frac{\alpha}{\beta}\right) < 1 \\ 1 & \left(\frac{\alpha}{\beta}\right) \geq 1 \end{cases} \tag{26}$$

We therefore expect that continuous time systems exhibit the same type of behaviors as the disretized versions above.

## 4 Conclusion

In this paper we have introduced a set of local controllers that are capable of controlling distributed systems that exhibit highly complicated behavior. These local controllers need only perturb local parameters and use local information about the state of the system. This eliminates the wiring and overhead needed for implementing a global or central controller. Through a statistical model, an analytical description of the dynamics of the controlled and uncontrolled degrees of freedom was obtained. We have also shown that there are critical values for the lattice couplings among the nonlinear elements beyond which the local controllers do not achieve the desired final state.

Two remarks concerning our results are in order. First, the failure to control the behavior of the system represented by the example of the lattice map Eq. 11, is due to groups of elements that stabilize their dynamics outside the regions in which controllers have an influence. These states then lead to rather "novel" patterns in the final state of the lattice that do not allow the whole system to reach the desired state. It is not certain however, how representative such "novel" final states are of continuous time systems, since a simple modification in the way the controllers were updated eliminated these patterns.

Second, we point out that the control of the system exemplified by the lattice map Eq. 4, can be better understood by resorting to a statistical model. Within this approach, uncontrolled lattice elements execute complicated orbits that tend to disrupt other elements in the lattice. Depending on the strength of the couplings and the dynamic ranges of the controllers, these uncontrolled regions either spread throughout the system or eventually disappear.

With the projected development of increasing smaller electromechanical devices, the issue of controlling vast distributed assemblies of nonlinear elements will surely become important. The current control engineer's toolbox is best equipped to deal with linear and low dimensional regimes, with the result that generalizations to higher dimensional systems are not obvious and can also be computationally expensive. We believe that the approach considered in this paper will become of practical importance as electromechanical systems become more integrated and complex.

This research was partially supported by ONR contract N00014–92–C-0046.

## References

[1] M. Mehregany. Microelectromechanical systems. *IEEE Circuits and Devices Magazine*, 9(4):14–22, 1993.

[2] F. Austin, M. J. Rossi, W. V. Nostrand A. Jameson, and J. Su. Active rib experiment for adaptive conformal wing. In *Third International Conference on Adaptive Structures*, pages 43–55, 1992.

[3] K.J. Moore, M. Noori, J. Wilson, and J.V. Dugan. The use of adaptive structures in reducing drag on underwater vehicles. In *Proceedings of the ADPA/AIAA/ASME/SPIE Conference on Active Materials and Adaptive Structures*, pages 331–334, 1991.

[4] L. Kleinrock. Distributed systems. *Communications of the ACM*, 28(11):14–27, 1985.

[5] B. Huberman and T. Hogg. The emergence of computational ecologies. In *1992 lectures in Complex Systems*, pages 185–205, 1991.

[6] Stanislas Leibler. Recent developements in the physics of fluctuating membranes. In L. Peliti, editor, *Biologically Inspired Physics*, pages 81–94. Plenum Press, 1989.

[7] F. Romeiras, C. Grebogi, E. Ott, and W.P. Dayawansa. Controlling chaotic dynamical systems. *Physica D*, 58:43–55, 1992.

[8] S. Watanabe and S.H. Strogatz. Integrability of a globally coupled oscillator arra. *Phys. Rev. Lett.*, 70(16):2391–2394, 1989.

[9] M. Silber, L. Fabiny, and K. Wiesenfeld. Stability results for in-phase and splay-phase states of solid-state laser arrays. *Journal of the Optical Society of America*, 10(6):1121–1129, 1993.

[10] A. T. Winfree. *The Geometry of Biological Time*. Springer-Verlag, New York, 1980.

[11] J.A. Sepulchre and A. Babloyantz. Controlling chaos in a network of oscillators. *Physical Review E*, 48(2):945–950, 1993.

[12] Rajarshi Roy, T.W. Murphy, T.D. Maier, Z. Gills, and E.R. Hunt. Dynamical control of a chaotic laser: Experimental stabilization of a globally coupled system. *Physical Review Letters*, 68(9):1259–1262, 1992.

[13] H. Gang and He Kaifen. Controlling chaos in systems described by partial differential equations. *Physical Review Letters*, 71(23):3794–3797, 1993.

[14] K. Kaneko. Pattern dynamics in spatiotemporal chaos. *Physica D*, 34:1–41, 1989.

[15] B.A. Huberman and E. Lumer. Dynamics of adaptive systems. *IEEE Trans. on Circ. Sys.*, 37:547–550, 1990.

[16] E. Ott, C. Grebogi, and J.A. Yorke. Controlling chaos. *Physics Review Letters*, 64:1196–1199, 1990.

[17] H. Gang and Q. Zhilin. Controlling spatiotemporal chaos in coupled map lattice systems. *Physics Review Letters*, 72(1):68–71, 1994.

[18] E.D. Lumer and G. Nicolis. Synchronous versus asynchronous dynamics in spatially distributed systems. *Unpublished*, 1993.